
%
%
%
\NPrefs
\def\define#1#2\par{\def#1{\Ref#1{#2}\edef#1{\noexpand\refmark{#1}}}}
\def\con#1#2\noc{\let\?=\Ref\let\<=\refmark\let\Ref=\REFS
         \let\refmark=\undefined#1\let\Ref=\REFSCON#2
         \let\Ref=\?\let\refmark=\<\refsend}

\define\V
G. Veneziano, Phys. Lett. {\bf B167}(1985) 388;~J. Maharana and G.
Veneziano, Phys. Lett. {\bf B169}(1986) 177.

\define\VM
J. Maharana and G. Veneziano, Nucl. Phys. {\bf B283}(1987) 126.

\define\EO
M. Evans and B. Ovrut, Phys. Rev. {\bf D41}(1990) 3149.

\define\GR
D. Gross, Phys. Rev. Lett. 60(1990).

\define\FR
E. Fradkin, Private communication.

\define\SZ
M. Saadi and B. Zwiebach, Ann. Phys. {\bf 192}(1989) 213;
T. Kugo, H. Kunitomo and K. Suehiro, Phys. Lett. {\bf B226} (1989) 48;
M. Kaku and J. Lykken, Phys. Rev. {\bf D3} (1988) 3067;
M. Kaku, preprint CCNY-HEP-89-6, Osaka-OU-HET 121;
C. Schubert, MIT preprint CTP 1977.

\define\KS
T. Kugo and K. Suehiro, Nucl. Phys.{\bf B337} (1990) 343.

\define\GS
D. Ghoshal and A. Sen, Tata preprint, TIFR/TH/91-47.

\define\AS
A. Sen, Phys. Lett. {\bf B241} (1990) 350, Nucl. Phys. {\bf B345} (1990)
551.

\define\MMV
J. Maharana, N. E. Mavromatos and G. Veneziano, unpublished.

\define\MMSO
S. Mukherji, S. Mukhi and A. Sen, Phys. Lett. {\bf B266} (1991) 337.

\define\IM
C. Imbimbo and S. Mukhi, Nucl. Phys. {\bf B364} (1991) 662.

\define\POL
A. M. Polyakov, Mod. Phy. Lett. {\bf A6} (1991) 635.

\define\MMST
S. Mukherji, S. Mukhi and A. Sen, Tata preprint, TIFR/TH/91-28, to appear
in Phys. Lett. B.

\define\KP
I. R. Klebanov and A. M. Polyakov, Mod. Phys. Lett. {\bf A6}(1991)3273.

\define\WIT
E. Witten, Institute for advanced study preprint, IASSNS-91/51

\define\EMN
J. Ellis, N. E. Mavromatos and D. V. Nanopoulos, Phys. Lett. {\bf B272}
(1991) 262.

\define\ASE
A. Sen, Tata preprint, TIFR/TH/90-49, to appear in Int. Jour. Mod. Phys.
{\bf A}.

{}~\hfill\vbox{\hbox{TIFR/TH/92-01}\hbox{IP/BBSR/91-54}\hbox{1991-92}}\break

\def\cm{c_0^{-}}
\def\lm{L_0^{-}}
\def\cp{c_0^{+}}
\def\bm{b_0^{-}}
\def\a1{{\alpha_{-1}}}
\def\a{{\alpha_{-1}}}
\def\a2{{\alpha_{-2}}}
\def\r{\rangle}
\def\l{\langle}
\def\tpmu{{\tilde p}_{\mu \nu \rho}}
\def\tdmu{{\tilde d}_{\mu \nu  \rho}}
\def\tmmu{{\tilde m}_{\mu \nu}}
\def\tnmu{{\tilde n}_{\mu \nu}}
\def\thmu{{\tilde h}_{\mu \nu \rho \sigma}}
\def\p{\partial }
\def \bp{\bar \partial}

\title{ GAUGE TRANSFORMATIONS IN STRING FIELD THEORY AND CANONICAL
TRANSFORMATIONS IN STRING THEORY}

\author{ Jnanadeva Maharana}
\address{ Institute of Physics\break Sachivalaya Marg, Bhubaneswar 751005,
India}
\author{ Sudipta Mukherji\foot{e-mail: mukherji@tifrvax}}
\address {Tata Institute of Fundamental Research \break Homi Bhabha Road,
Bombay 400005, India}

\abstract

We study how canonical transfomations in first quantized string theory can
be understood as gauge transformations in string field theory. We
establish this fact by working out some examples. As a by product, we
could identify some of the fields appearing in string field theory
with their counterparts in the $\sigma$-model.

\submit{Physics Letters B}
\endpage

{\noindent {\bf {1. Introduction}}

The string theory is endowed with a rich symmetry structure. The ward
identities, derived thorugh the introduction of canonical transformations,
unravel the hidden symmetries of string theory. It was shown \V\VM~ that
ward identities associated with local symmetries such as general
coordinate invariance, abelian gauge invariance (associated with the
antisymmetric tensor field) and nonabelian gauge invariance (in the case
of compactified bosonic strings), could be derived in the Hamiltonian path
integral formalism. Furthermore, it was suggested in ref.\VM~(see also
\EO) that massive modes of the string might possess local symmetries and
the existance of such symmetries might be manifested through the
appropriate new ward identities. Indeed, it has been argued \GR\FR~that
all the string states are gauge particles and most of these gauge
symmetries are broken spontaneously leaving only the familiar local
symmetries as exact invariances of the string theory. Recently, it has
been proposed that quantum coherence is maintained by two dimensional
target space black holes in string theory due to the existance of infinite
set of conserved quantum numbers. It has been argued \EMN~that there is an
intimate connection between these conserved currents and the
stringy(higher) symmetries.

It is natural to expect that the string field theory is the appropriate
setting to provide a deepar understanding of the underlying(hidden)
symmetry structure in string theory.
Now that we have a consistent covariant quantum field theory of closed
bosonic string\SZ\KS, we can start asking these questions.
This theory is characterized by infinite number of
fields and  it possesses
non-linear gauge invariance.
It has been shown that the general coordinate transformations
arise as a particular combination of gauge transformations in this string
field theory and metric can be related to the infinite component string
field through suitable functional relation\GS.

In this letter we will concentrate on some specific gauge transformations
in string field theory
involving higher level states and will identify those as canonical
transformations of the target space
coordinate $X^{\mu}$ appearing in first quantized string theory.
In the $\sigma$-model language these
transformations generate symmetries in the
space  of the couplings
which appear in the $\sigma$-model
action. Throughout this letter we consider only
the linearized version of gauge transformations in string field theory and
at the end we discuss how this can be extended to take care of full
non-linear gauge transformations following the prescription of\GS.

This
letter is organized as follows. First we set our notations through stating
some known results about gauge invariance in string field theory. In
sec.3 we focus on some particular gauge transformations. In sec.4 we identify
these gauge transformations
with some canonical transformations on $X^{\mu}$ of the first
quantized string theory. This automatically allows us to
identify the fields appearing in string field theory with the coupling
constants
appearing in the $\sigma$-model describing
the first quantized string
theory. We conclude in sec.5 with some speculations and
possible extensions of the present work.

{\noindent{\bf 2. Gauge invariance in string field theory}}

Here we set our notations by stating some known facts about
the gauge symmetries in non-polynomial closed string field
theory. The configuration space of the string field $\Psi $ is generally
taken to be a subspace of the full Hilbert space ${\cal H}$ of ghost number
three  and is annihilated by $\cm $ and $\lm $. The string field theory
action involving $\Psi $ has the form
$$
S(\Psi) = {1\over 2}\l\Psi| Q_B \bm |\Psi\r + \sum _{N=3}^{\infty}
{g^{N-2}\over {N!}} \{\Psi^N\}
\eqn\one
$$
Here $Q_B$ is the BRST charge of the first quantized theory and $\{ \}$
has been defined in\KS\AS.
The string field $\Psi$ can be expanded in some basis as $\bm|\Psi\r =
\sum_{r} \psi_{r} |\Phi_{r}\r$ and in terms of component fields the
action becomes
$$
S(\Psi) = \sum_{N=2}^{\infty} {1\over N} A_{r_1....r_N}^{(N)} \psi_{r_1}..
. \psi_{r_N}
\eqn\two
$$
$A^{(N)}_{r_1...r_N}$ are constants which are computable using
the techniques of conformal field theory. The classical equations of
motion derived from \one~have the form
$$
Q_B \bm |\Psi\r + \sum_{N=3}^{\infty}{g^{N-2}\over{(N-1)!}}[\Psi^{N-1}] =
0
\eqn\three
$$
where $\l A_1|[ A_2 ...A_N]\r = (-1)^{n_1+1} \{ A_1...A_N\}$ with $n_1$
being the ghost number of $A_1$.

The action is invariant under the infinitesimal gauge transformation
$$
\delta (\bm \Psi) = Q_B \bm \Lambda + \sum_{n=3}^{\infty}
{g^{N-2}\over{(N-2)!}}[\Psi^{N-2}\Lambda ]
\eqn\four
$$
Here $\Lambda$, the infinitesimal gauge transformation parameter, is a
state with ghost number 2 and is annihilated by $\cm$ and $\lm$. As
before $\Lambda$ can be expanded in components as $ \bm |\Lambda\r =
\sum_{\alpha} \lambda_{\alpha} |\eta_{\alpha}\r$.
Then \four~can be written as
$$
\delta \psi_s = \sum_{N=2}^{\infty} B_{s\alpha
r_1...r_{N-2}}^{(N)}\lambda_{\alpha}
\psi_{r_1}... \psi_{r_{N-2}}
\eqn\five
$$
where $B_{r_1...r_{N-2}}^{(N)}$ are computable constants.

Notice that the infinitesimal gauge transformations contain linear as well
as
non-linear parts. To linear order all the field dependent parts drop out and
hence gauge transformations change the fields by adding BRST exact
states to it.

{\noindent{\bf{3. Some specific off-shell linearised gauge transformations}}}

In this section we will concentrate on some specific gauge transformations
which essentially amount to choosing some specific $\Lambda$ in
\four.~The idea is to choose $\Lambda$ such a way that it can be identified
as
a canonical transformation in the
$\sigma$-model constructed from
the first quantized string theory. This will allow us to identify the
off-shell field configurations of string field theory with the coupling
constants
appearing in the $\sigma$-model.

For that purpose let us choose the gauge transformation parameter to be
$$
\bm|\Lambda\r = \int d^{26} k \Gamma_{\mu\{\nu\rho\}}(k)
({\alpha_{-1}} ^{\mu}c_1 -
{\bar{\alpha}}_{-1}^{\mu} {\bar c_1}) {\alpha_{-1}}^{\nu}
{\bar {\alpha}}_{-1}^{\rho}|k\r
\eqn\five
$$
with $\Gamma_{\mu\{\nu\rho\}}$ being symmetric under $\nu$ and $\rho$
interchange.
Notice that $|\Lambda\r = \cm\bm |\Lambda\r$
satisfies all the required properties
of a well defined state in the Hilbert space,
namely, it has the
correct ghost number and is annihilated by $\cm$ and $\lm$. We now define the
string field configuration involving rank 2, 3, and 4 tensor fields as
$$\eqalign{
\bm|\Psi\r =& \int d^{26}k[{\tilde p}_{\mu \nu \rho}\cp
(\alpha_{-1}^{\mu}\alpha_{-1}^{\nu}
{\bar{\alpha}}_{-1}^{\rho}c_1 - {\bar {\alpha}}_{-1}^{\mu}
{\bar {\alpha}}_{-1}^{\nu} \alpha_{-1}^{\rho}
\bar{c_1})\cr
& + \tdmu c_1 {\bar {c}_1}(\alpha_{-1}^{\mu}
\alpha_{-1}^{\nu}{\bar{\alpha}}_{-2}^{\rho} + {\bar
{\alpha}}_{-1}^{\mu}{\bar {\alpha}}_{-1}^{\nu}\alpha_{-2}^{\rho})\cr
&+\tmmu (\alpha_{-1}^{\mu}
{\bar {\alpha}}_{-1}^{\nu}c_{-1}c_{1} - {\bar {\alpha}}_{-1}^{\mu}
\alpha_{-1}^{\nu} {\bar {c}}_{-1}{\bar {c}}_1)\cr
&+\tnmu (\alpha_{-1}^{\mu}
\alpha_{-1}^{\nu}c_1
{\bar {c}}_{-1} -
{\bar {\alpha}}_{-1}^{\mu}{\bar {\alpha}}_{-1}^{\nu}{\bar {c}}_1 c_{-1})\cr
&+\thmu c_1{\bar{c}}_1 ( \alpha_{-1}^{\mu}
\alpha_{-1}^{\nu} {\bar {\alpha}}_{-1}^{\rho} {
\bar {\alpha}}_{-1}^{\sigma} +
{\bar {\alpha}}_{-1}^{\mu}{\bar {\alpha}}_{-1}^{\nu} \alpha_{-1}^{\rho}
\alpha_{-1}^{\sigma})\cr
&+......]|k\r .\cr}
\eqn\seven
$$
Here $\tpmu, \tdmu, \tmmu, \tnmu, \thmu$ are the fields of different
tensorial ranks
(all of them are not necessarily physical)
and ... contains all other
fields which are not important for our present discussion.

As mentioned earlier, at the linearized level the gauge transformations in
string field theory reduce to
$$
\delta (\bm |\Psi\r) = Q_B \bm |\Lambda\r.
\eqn\eight
$$
The BRST charge has the following decomposition in terms of matter and
ghost Virasoro generator.
$$
Q_B = \sum_{-\infty}^{\infty} : (L_{-m}^{matter} + {1\over
2}L_{-m}^{ghost}) c_m: + c.c.
\eqn\nine
$$
It is now straight forward to calculate $Q_B\bm|\Lambda\r$ with $\Lambda$
given in \five. Computing that we get
$$\eqalign{
Q_B\bm |\Lambda\r =& [{k^2\over \sqrt 2} \Gamma_{\mu \{\nu \rho\}}
            \alpha_{-1}^{\mu}\alpha_{-1}^\nu {\bar {\alpha}}_{-1}^\rho\cp c_1
              -{k^2\over \sqrt 2}\Gamma_{\mu\{\nu \rho\}}{\bar
{\alpha}}_{-1}^ \mu \alpha_{-1}^\nu {\bar {\alpha}}_{-1}^\rho \cp {\bar
c}_{1}\cr
& + {k^\mu \Gamma_{\mu\{\nu \rho\}} \alpha_{-1}^{\nu}{\bar
{\alpha}}_{-1}^{\rho} c_{-1}c_{1} +  k^\mu \Gamma_{\mu\{\nu\rho\}}
\alpha_{-1}^{\mu}{\bar {\alpha}}_{-1}^{\rho} c_{-1}c_{1}}\cr
& -{ k^\rho \Gamma_{\mu\{\nu \rho\}}\alpha_{-1}^\mu
\alpha_{-1}^\nu c_1{\bar {c}}_{-1} - k^\rho \Gamma_{\mu\{\nu \rho\}}
\alpha_{-1}^\mu \alpha_{-1}^\nu {\bar c_{-1}}{\bar c_1}}\cr
& - k^\mu \Gamma_{\mu\{\nu \rho\}}
\alpha_{-1}^\nu{\bar{\alpha}}_{-1}^\rho {\bar c_{-1}}{\bar c_1}
 - k^\nu \Gamma_{\mu\{\nu \rho\}}{\bar{\alpha}}_{-1}^\mu {\bar
{\alpha}}_{-1}^\rho c_1{\bar {c}}_{-1}\cr
& -\Gamma_{\mu\{\nu \rho\}}\alpha_{-1}^\mu \alpha_{-1}^\nu
{\bar{\alpha}}_{-2}^\rho c_1\bar c_{-1}
 -\Gamma_{\mu\{\nu \rho\}}{\bar{\alpha}}_{-1}^\mu \alpha_{-2}^\nu
{\bar{\alpha}}_{-1}^\rho c_1\bar c_{-1}\cr
& - k_\gamma\Gamma_{\mu \{\nu \rho\}}
\alpha_{-1}^{\mu}\alpha_{-1}^\nu {\bar
{\alpha}}_{-1}^\rho{\bar{\alpha}}_{-1}^\gamma  c_1 \bar c_1
 - k_\gamma\Gamma_{\mu \{\nu \rho\}}
\alpha_{-1}^{\nu}\alpha_{-1}^\gamma {\bar
{\alpha}}_{-1}^\rho{\bar{\alpha}}_{-1}^\mu  c_1 \bar c_1]|k\r.\cr}
\eqn\ten
$$
Using \eight, \ten~in \seven~we find the gauge transformed string field
now has the form in terms of components as
$$\eqalign{
\bm|\Psi\r = & \int d^{26}k[(\tpmu +{k^2\over \sqrt 2}\Gamma_{\mu\{\nu\rho\}}
)\cp
(\alpha_{-1}^{\mu}\alpha_{-1}^{\nu}
{\bar{\alpha}}_{-1}^{\rho}c_1 - {\bar {\alpha}}_{-1}^{\mu}
{\bar {\alpha}}_{-1}^{\nu} \alpha_{-1}^{\rho}
\bar{c_1})\cr
& + (\tmmu + k^\rho \Gamma_{\rho\{\mu\nu\}} +k^\rho \Gamma_{\mu\{\rho
\nu\}})( \alpha_{-1}^\mu
{\bar {\alpha}}_{-1}^{\nu}c_{-1}c_{1} - {\bar {\alpha}}_{-1}^{\mu}
\alpha_{-1}^{\nu} {\bar {c}}_{-1}{\bar {c}}_1)\cr
& + (\tnmu - k^\rho \Gamma_{\mu\{\nu\rho\}}) (\alpha_{-1}^{\mu}
\alpha_{-1}^{\nu}c_1
{\bar {c}}_{-1} -{\bar {\alpha}}_{-1}^{\mu}
{\bar {\alpha}}_{-1}^{\nu}{\bar {c}}_1 c_{-1})\cr
& + (\thmu + k_{\sigma}\Gamma_{\mu\{\nu \rho\}})
c_1{\bar{c}}_1 ( \alpha_{-1}^{\mu}
\alpha_{-1}^{\nu} {\bar {\alpha}}_{-1}^{\rho} {
\bar {\alpha}}_{-1}^{\sigma} +
{\bar {\alpha}}_{-1}^{\mu}{\bar {\alpha}}_{-1}^{\nu} \alpha_{-1}^{\rho}
\alpha_{-1}^{\sigma})\cr
& + ( \tdmu + \Gamma_{\mu\{\nu\rho\}})
c_1 {\bar {c}_1}(\alpha_{-1}^{\mu}
\alpha_{-1}^{\nu}{\bar{\alpha}}_{-2}^{\rho} + {\bar
{\alpha_{-1}}}^{\mu}{\bar {\alpha}}_{-1}^{\nu}\alpha_{-2}^{\rho})+....
]|k\r.\cr}
\eqn\eleven
$$
Hence the corresponding fields transform as
$$\eqalign{
& \delta \tpmu = {k^2\over \sqrt 2} \Gamma_{\mu\{\nu\rho\}}\cr
& \delta \tmmu =k^\rho \Gamma_{\rho\{\mu\nu\}} + k^\rho
\Gamma_{\mu\{\rho\nu\}}\cr
& \delta \tnmu = - k^\rho \Gamma_{\mu\{\nu\rho\}}\cr
& \delta \tdmu = \Gamma_{\mu\{\nu\rho\}}\cr
&\delta \thmu = k_\sigma \Gamma_{\mu\{\nu\rho\}}.\cr}
\eqn\twelve
$$
Let us now redefine the fields as
$$\eqalign{
& p_{\mu \nu \rho} = \tpmu -{k^\sigma\over \sqrt 2}\thmu\cr
& m_{\mu\nu} = \tmmu - k^\rho {\tilde d}_{\rho\mu\nu} - k^\rho {\tilde
d}_{\mu\rho\nu}\cr
& n_{\mu\nu} = \tnmu + k^\rho {\tilde d}_{\mu\nu\rho}\cr
& d_{\mu\nu\rho} = \tdmu\cr
& h_{\mu\nu\rho\sigma} = \thmu.\cr}
\eqn\thirteen
$$
Tramsformation laws of all the fields now take the form
$$\eqalign{
& \delta p_{\mu\nu\rho} = 0\cr
& \delta m_{\mu\nu} = 0\cr
& \delta n_{\mu\nu} = 0\cr
& \delta d_{\mu\nu\rho} = \Gamma_{\mu\{\nu\rho\}}\cr
& \delta h_{\mu\nu\rho\sigma} = k_\sigma \Gamma_{\mu\{\nu\rho\}}.\cr}
\eqn\fourteen
$$

{}From \three~we see that the linearized equation of motion in string field
theory is\break
$Q_B\bm |\Psi\r = 0$. Using this
we can easily find out the equations of motion for each component fields
appearing in the expansion of $\Psi$. Equations corresponding to
the fields $p_{\mu\nu\rho}$, $m_{\mu\nu}$ and $n_{\mu\nu}$ are purely
algebraic in nature and can be identified as constraint equations. Hence
we can set these fields to be zero at least to this order. In that case
\seven ~reduces to
$$\eqalign{
\bm|\Psi\r =& \int d^{26}k[d_{\mu\nu\rho}c_1{\bar c}_1(\alpha_{-1}^\mu
\alpha_{-1}^\nu {\bar \alpha}_{-2}^\rho + {\bar \alpha}_{-1}^\mu
{\bar \alpha}_{-1}^{\nu} \alpha_{-2}^{\rho})\cr
& + h_{\mu\nu\rho\sigma}c_{1}{\bar c}_1(\alpha_{-1}^\mu \alpha_{-1}^\nu
{\bar \alpha}_{-1}^\rho {\bar \alpha}_{-1}^\sigma + {\bar\alpha}_{-1}^\mu
{\bar\alpha}_{-1}^\nu \alpha_{-1}^\rho \alpha_{-1}^\sigma)\cr
& +.....]|k\r.\cr}
\eqn\fifteen
$$
Notice that here we have replaced $\tdmu$ and $\thmu$ in
terms of $d_{\mu\nu\rho}$
and $h_{\mu\nu\rho\sigma}$ using \thirteen. Substituting \fifteen~in
the quadratic part of the action given in \one~it is now trivial
to write down the action in terms
of component fields $d_{\mu\nu\rho}$ and
$h_{\mu\nu\rho\sigma}$.

Before ending this section we would like to mention that instead of
choosing \five~as gauge transformation parameter, we can as well take
$$
\bm |\Lambda\r = \int d^{26}k {\Gamma^\prime }_{\mu[\nu\rho]}(k)
  (\alpha_{-1}^\mu c_{1} - {\bar\alpha}_{-1}{\bar c}_1)
  \alpha_{-1}^\nu {\bar \alpha}_{-1}^{\rho}|k\r.
\eqn\sixteen
$$
Here ${\Gamma^\prime}_{\mu[\nu\rho]}$ is antisymmetric under interchange
of indeces $\nu$ and $\rho$.
Following the same prescription as before we can find out how the
relavent fields change under this gauge transformation.

\noindent{\bf 4. Canonical transformations in first quantized string theory}

In first quantized string theory one writes down the action in terms
of the target space coordinate $X^\mu$ in the presence of background
excitations of the string. One can as well write the action
in terms of canonically transformed variable ${X^\prime}^\mu$ and its
conjugate momentum ${P^\prime}^\mu$. In
$\sigma$-model laguage, these canonical transformations genarate
symmetry in the parameter space of the theory. In this section
we will concentrate
on some particular canonical transformations and will identify them as
the gauge transformations in string field theory discussed in the previous
section.

Let us make the following canonical transformation\MMV
$$
{X^\prime }^\mu = X^\mu + \p X^\nu \bp X^\sigma \zeta_{\{\nu\sigma\}}^\mu .
\eqn \seventeen
$$
We can find out how the parameters in the $\sigma$-model change under this
transformation. We take the $\sigma$-model action as follows:
$$\eqalign{
A = & \int d^2 z [ T(X) + (\p X^\mu \bp X^\rho + \p X^\rho \bp X^\mu)
G_{\mu\rho}(X)\cr
& + D_{\mu\nu\rho}(X)(\p X^\mu\p X^\nu \bp^2 X^\rho
+\p^2 X^\mu \bp X^\nu \bp X^\rho)\cr
& + H_{\mu\nu\rho\sigma}(X)(\p X^\mu \p X^\nu \bp X^\rho \bp X^\sigma
+ \bp X^\mu \bp X^\nu \p X^\rho \p X^\sigma)\cr
& +......].\cr}
\eqn\eighteen
$$
Under \seventeen
$$
\delta T(X) = T(X)_{,\mu} \p X^\nu \bp X^\sigma
\zeta_{\{\nu\sigma\}}^{\mu}
\eqn\nineteen
$$
and
$$\eqalign{
\delta [G_{\mu\rho}(& \p X^\mu \bp X^\rho +  \p X^\rho \bp X^\mu)]\cr
& =  (\p X^\nu \bp X^\sigma \p X^\lambda \bp X^\rho
\zeta_{\{\nu\sigma\},\lambda }^\mu + \p^2 X^\nu \bp X^\sigma \bp X^\rho
\zeta_{\{\nu\sigma\}}^\mu)
G_{\mu\rho}\cr
& + ( \p X^\mu \p X^\nu \bp X^\sigma \bp X^\lambda
\zeta_{\{\nu\sigma\},\lambda}^\rho + \bp^2 X^\sigma \p X^\nu \p X^\mu
\zeta_{\{\nu\sigma\}}^\rho)
G_{\mu\rho}\cr
& + ( \p X^\nu \p X^\lambda \bp X^\sigma \bp X^\mu
\zeta_{\{\nu\sigma\},\lambda}^\rho + \p^2 X^\nu \bp X^\sigma \bp X^\mu
\zeta_{\{\nu\sigma\}}^\rho)
G_{\mu\rho}\cr
& + ( \p X^\nu \p X^\rho \bp X^\sigma \bp X^\lambda
\zeta_{\{\nu\sigma\},\lambda}^\mu + \bp^2 X^\sigma \p X^\rho \bp X^\nu
\zeta_{\{\nu\sigma\}}^\mu)
G_{\mu\rho}\cr
& + ( \p X^\mu \p X^\nu \bp X^\sigma \bp X^\rho
\zeta_{\{\nu\sigma\}}^\lambda G_{\mu\rho,\lambda}
 + \p X^\rho \p X^\nu \bp X^\mu \bp X^\sigma
\zeta_{\{\nu\sigma\}}^\lambda G_{\mu\rho,\lambda}).\cr}
\eqn\twenty
$$
Comparing \eighteen~and \nineteen~we see that under \seventeen~~the
transformed graviton becomes
$$
G^\prime_{\mu\nu}(X)
= G_{\mu\nu}(X) + \zeta _{\{\mu\nu\}}^\alpha T_{,\alpha}(X) .
\eqn\twentyone
$$
Similarly comparing \twenty~and \eighteen~we see that the three and four
rank tesors are going to mix with graviton as
$$\eqalign{
& H_{\mu\nu\rho\sigma}^\prime (X) = H_{\mu\nu\rho\sigma}(X) + 2
\zeta_{\{\mu\rho\},\nu}^\lambda G_{\sigma\lambda}(X)
+ \zeta_{\nu\sigma}^{\lambda} G_{\mu\rho,\lambda}(X)\cr
& D_{\mu\nu\rho}^\prime(X) =  D_{\mu\nu\rho}(X)
+ 2 \zeta_{\mu\nu}^\lambda G_{\rho\lambda}(X).\cr}
\eqn\twentytwo
$$
If we expand now $G_{\mu\nu}$ around flat background as $G_{\mu\nu} =
\eta_{\mu\nu} + h_{\mu\nu}$, we get from \twentytwo~(keeping only those
terms which are lowest order in fields)
$$
H_{\mu\nu\rho\sigma}^\prime(X) = H_{\mu\nu\rho\sigma}(X) + 2
\zeta_{\sigma\{\mu\rho\},\nu}
\eqn\twentythree
$$
and
$$
D_{\mu\nu\rho}^\prime(X) =
D_{\mu\nu\rho}(X) + 2 \zeta_{\rho\{\mu\nu\}}.
\eqn\twentyfour
$$
Graviton remains unchanged to this order. Now looking at the transformation
laws of various fields appearing in string field theory under the gauge
transformation discussed earlier ( see \thirteen), it is easy to see
the canonical transformation \seventeen~does the same job in the first
quantized theory if we make the following idetification:
$$\eqalign{
& d_{\mu\nu\rho} = D_{\mu\nu\rho}\cr
& h_{\mu\nu\rho\sigma} = H_{\mu\nu\rho\sigma}\cr}
\eqn\twentyfive
$$
and
$$
-{i\over 2} \int d^{26} k \Gamma_{\mu\{\nu\rho\}}(k)e^{ikx} =
\zeta_{\mu\{\nu\rho\}} (x).
\eqn\twentysix
$$
In other words,
above identifications
allow us to relate a particular canonical transformation
in the $\sigma$-model with a gauge
transformation in the string field theory.

Finally, notice that instead of \seventeen~we can choose a
transformation
$$
{X^\prime}^\mu = X^\mu + \p X^\nu \bp X^\rho
{\zeta^\prime}_{[\nu\rho]}^\mu .
\eqn\twentyseven
$$
In the language of string field theory this would presumably be a gauge
transformation with parameter given in \sixteen~if we make relevant field
identification though we haven't checked that explicitly.

{\noindent{\bf 5. Conclusion}}

To summarize, we have shown how canonical transformations in first
quantized string theory can be understood as off-shell gauge
transformations of some underlying string field theory. We could show that
by taking specific examples and working at the linearized level. As a by
product, we could make identification of some of the fields appearing in
string field theory with their counterparts in the $\sigma$-model.

Notice that the whole analysis can be performed
 for one dimensional strings coupled to gravity theories
with very little modification. It
is well known that these theories have a two dimensional target space
interpretation. Hence the general couplings are functions of two
variables, one being the conformal mode of the two dimensional target
space metric. It has been noticed earlier that these theories possess
infinite sequence of discrete states, besides tachyon\POL\MMSO.
Hence one can write
down a $\sigma$-model action involving these states. Consequently the
canonical transformations are now suitable redefinitions of the two
dimesional target space coordinates $X^\mu (\mu=1,2)$. On the other hand,
one can construct a string field theory action\MMST\IM~for these theories.
To make the total central charge 26, one needs to introduce background
charge for such non-critical string theories. Hence the Virasoro
generators and the BRST charge gets modified. But the structure of the
gauge transformations remain same as \four. Now if we take the canonical
transformations \seventeen~for these 2-dim. theories, it will certainly
have corresponding gauge transformation parameter in string field theory
similar to \five. It has been conjectured \EMN~that such canonical
transformations are responsible for $W_{\infty}$ symmetries for 2-dim.
string theories (see also \KP\WIT).
Once one finds the corresponding gauge transformation
parameters in string field theory, it will be interesting to check that
explicitly and analyse the consequences following \ASE.

Throughout this letter we have considered only the linearized gauge
transformations and so we could compare only the quadratic part of the
string field theory action
with the corresponding part in the $\sigma$-model
action. If we want to go beyond the linear level, changes of the component
fields in \seven~will be completely non-linear. Similarly, for the case of
canonical transformation in $\sigma$-model action, in \twentythree~and
\twentyfour~we need to keep higher order field dependent terms. It would
be interesting to investigate whether similar kind of identifications as
\twentyfive~can be made. This can be done systemetically following the
algorithm developed in \GS.

We hope to come back to these issues in future.

\noindent {\bf Acknowledgement:}
We are extremely grateful to Ashoke Sen for several discussions
at all stages of this work and for his comments on the manuscript.
We would also like to thank Sumit Das for interesting discussions.

\refout

\end